\documentclass[aps,prd,reprint,groupedaddress,amsmath,amssymb,showpacs,
floatfix]{revtex4-1}
\usepackage{graphicx}
\usepackage{dcolumn}
\usepackage{bm}
\begin{document}
\title{Light deflection in Kerr field for off-equatorial source}
\author{Sarani Chakraborty}
\email[]{sarani.chakraborty.phy@gmail.com}
\affiliation{Department of Physics, Assam University, Silchar-788011, Assam, India.}
\author{A. K. Sen}
\email[]{asokesen@yahoo.com}
\affiliation{Department of Physics, Assam University, Silchar-788011, Assam, India.}
\date{\today}
\begin{abstract}
Deflection angle for a light ray travelling in the equatorial plane of a rotating Kerr mass has been calculated in past by various investigators. Considering the light ray to be travelling only slightly above the equatorial plane, calculations have been made in the present paper for such a ray for its deflection angle. We calculate deflection angles for the light ray at various heights, which are small compared to the impact parameter and derive corresponding analytical expressions for deflection angle.
\end{abstract}

\maketitle

\section{INTRODUCTION}
In general relativity, Schwarzschild line element gives the solution of Einstein's field equation for the exterior of an uncharged, static mass [1]. Bending angle for Schwarzschild field was calculated by many authors up to second and higher order terms. Keeton et al. [2] calculated the light deflection angle for Schwarzschild black hole in weak field approximation. Iyer and Petters [3] calculated light deflection angle for strong field conditions, which under weak field approximation is in exact match with that of Keeton et al. [2].
\\The solution of Einstein's field equation which describes the exterior of an uncharged, rotating mass was obtained by Kerr in the year of 1965 [4]. According to this line element, the frame dragging an unusual prediction of the general relativity is exhibited by such rotating masses. The prediction of this effect is that, all objects coming close to a rotating mass would be entrained to participate in its rotation, not because of any applied force or torque that can be felt, but because of the curvature of space time associated with the rotating mass. At close enough distance all objects even light must rotate with the mass. So the light deflection angles for Schwarzschild and the Kerr geometries are not the same. There are some extra terms, which arise due to the effect of rotation.
\\Bending angle for Kerr field in equatorial plane (i.e $\theta=\frac{\pi}{2}$) was calculated by Iyer et al. [5, 6]. According to their result, deflection produced in presence of a rotating black hole explicitly depends on direction of motion of the light. If the light ray is moving in the direction of rotation (called the proggrade orbit), then the deflection angle is higher than that for the zero rotation Schwarzschild field. If the light ray is moving opposite to the direction of rotation (called the retrograde orbit), the bending angle will be smaller than that for Schwarzschild field. Dubey and Sen [7] had used Kerr geometry, to show how the gravitational redshifts are affected as photon is emitted at various latitudes of rotating mass. Kraniotis [8] derived the analytic solutions for the lens equations in terms of Appell and Lauricella hypergeometric functions and the Weierstra${\ss}$ modular form for a Kerr mass. All the above mentioned calculations (for Schwarzschild and Kerr) were done using the null geodesic for light ray.
\\On the other hand, another technique has been developing by some authors called effective refractive index of material medium. Here the effect of gravitation is seen as the change in the refractive index of the medium through which light is travelling. In the year 1958, Balaz [9] used this method to calculate the effect of gravity due to a rotating mass on the direction of polarization vector of an electromagentic wave. Atkinson [10] used this method to study the trajectory of light ray near a very massive, static and spherically symmetric star [10]. Fischbach and Freeman [11] derived the second order term of gravitational deflection using the same method. Sen [12] and Roy and Sen [13] calculated the light deflection angles for Schwarzschild and Kerr masses using this method.
\\All these above mentioned works for Kerr field were done considering the light ray contained in the equatorial plane. But there are authors who have worked on non-equatorial plane also. Bozza [14] obtained the lensing formula, and calculated the relativistic image position for a light ray trajectory close to equatorial plane of a Kerr black hole. However, he did not calculate the light deflection angle explicitly for the non-equatorial plane. Aazami et al. [15, 16] worked on quasi-equatorial regime of Kerr black hole and calculated the two components of light bending angle, along the direction of equatorial plane and perpendicular to the equatorial plane of a Kerr black hole. Their expression for light deflection angle along the direction of equatorial plane is in exact match with the result of Iyer et al. [6].
\\For a static charged gravitational field, the solution of Einstein field equation was first given by Reissner [17] and  Nordstr$\ddot{o}$m [18] independently, known as  Reissner-Nordstr$\ddot{o}$m solution. Eiroa et al [19] calculated the light deflection angle for this line element and showed that charge itself has some effect on curvature of space-time. Vibhadra et al. [20] obtained the light bending angle for Janis-Newman-Winicour space-time (another solution for static, charge mass) and it was shown that for zero static charge, it reduces to Schwarzschild bending angle. The exact solution of Einstein's field equation for a rotating charged mass was found by Newman et. al [21, 22] in the year 1967, which is known as Kerr-Newman metric. Chakraborty and Sen [23] have recently obtained the light deflection angle for a charged, rotating mass in the equatorial plane. Under the condition of zero charge, their expression for deflection angle reduces exactly to that of Iyer et al. [6].
\\In this paper we consider an uncharged, rotating Kerr mass where the light ray is coming from an off-equatorial source i.e. the source at infinity is not situated on the equatorial plane, but it is placed at a very small height $l$ above the equatorial plane (Fig 1). For a rotating body, rotation drags the trajectory of light ray towards the rotation plane i.e. the equatorial plane where the rotation has its maximum effect.
\\In this paper we consider a slowly rotating, massive gravitational body, where gravity dominates over the rotation. So the light ray remains on the gravitational plane i.e. all over its trajectory light ray maintains a height $l$ above the equatorial plane. Direct consequence of this assumption is that the latitude angle $\theta$ will be a constant. Strictly speaking as the light ray undergoes deflection the
height will not maintained constant all over its trajectory. However, for the ease of calculation we
assumed the height to be constant and such approached have been also adopted by other authors
in past. With this assumption we calculate in the present work the off-equatorial light deflection angle for a slowly rotating body.
\begin{widetext}
\begin{figure}
\includegraphics[width=13cm]{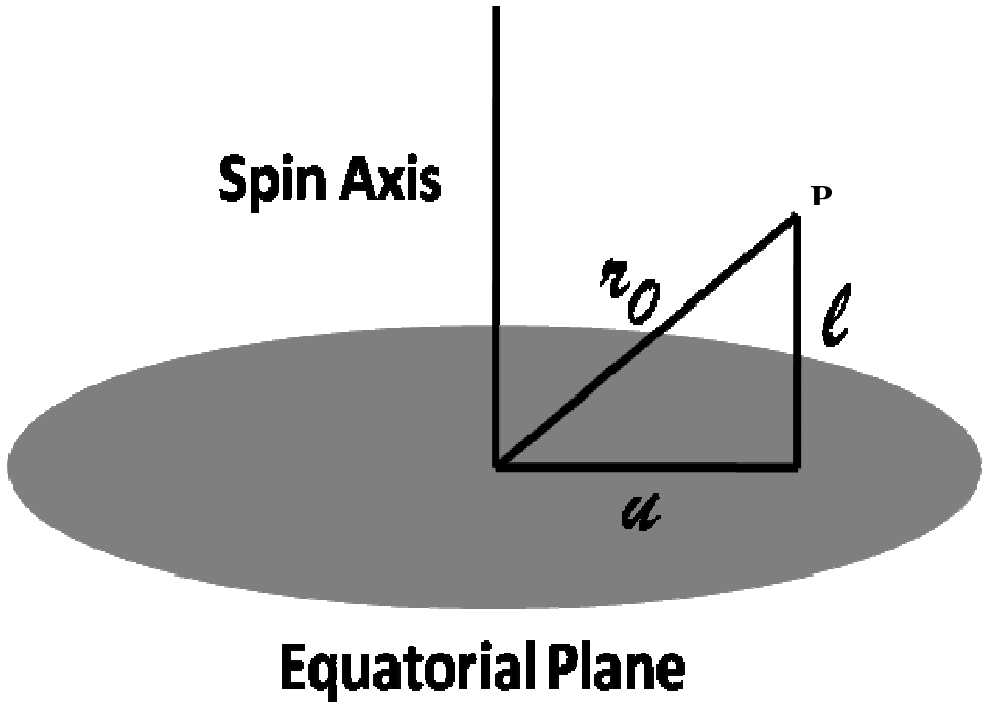}
\caption{Geometry of the equatorial plane of a rotating mass, where the light ray is passing through the point P, perpendicular to the plane of paper. Here, $l$ is the height of the light source above the equatorial plane. $u$ is the projected impact parameter on the equatorial plane and $r_{0}$ is the closest approach of the light ray. \label{f1}}
\end{figure}
\end{widetext}
\section{NULL GEODESIC IN KERR FIELD}

The Kerr line element in Boyer-Lindquist co-ordinate ($r,\theta,\phi,t$) can be written as [24, page 313]

\begin{equation}
 ds^{2}=(1-\frac{2mr}{\rho^{2}})c^{2}dt^{2}-\frac{\rho^{2}}{\Delta}dr^{2}-\rho^{2}d\theta^{2}-(r^{2}+a^{2}+$$
 $$\frac{2mra^{2}\sin^{2}\theta}{\rho^{2}})\sin^{2}\theta d\phi^{2}+\frac{4mra\sin^{2}\theta }{\rho^{2}}cdtd\phi
\end{equation}
where
\begin{equation}
 \rho=r^{2}+a^{2}\cos^{2}\theta
\end{equation}
\begin{equation}
 \Delta=a^{2}+r^{2}-2mr
\end{equation}
and $m=\frac{GM}{c^{2}}$ and $a=\frac{J}{cM}$, further c is the velocity of light in free space, M is the mass of the gravitating body, G is the gravitational constant and J is the angular momentum of the gravitating body.
Lagrangian ($\mathcal{L}$) can be written as, [24, page 96]
\begin{equation}
 \mathcal{L}=g_{\mu\nu}\dot{x}^{\mu}\dot{x}^{\nu}
\end{equation}
where the dot indicates derivative with respect to some affine parameter. So the lagrangian ($\mathcal{L}$) can be written as [24],
\begin{equation}
2\mathcal{L}= (1-\frac{2mr}{\rho^{2}})c^{2}\dot{t}^{2}+\frac{4amr\sin^{2}\theta}{\rho^{2}}c\dot{t}\dot{\phi}-\frac{\rho^{2}}{\Delta}\dot{r^{2}}-\rho^{2}\dot{\theta^{2}}- (r^{2}+a^{2}$$$$+\frac{2a^{2}mr\sin^{2}\theta}{\rho^{2}})\sin^{2}\theta\dot{\phi^{2}}
\end{equation}
From this equation, the energy (E) and angular momentum in the direction of rotation $(L_{z})$ can be written as [24, page 346]:
\begin{equation}
E=P_{t}=\frac{\partial \mathcal{L}}{\partial (ct)}=(1-\frac{2Mr}{\rho^{2}})c\dot{t}+\frac{2amr\sin^{2}\theta}{\rho^{2}}\dot{\phi}
\end{equation}
\begin{equation}
 L_{Z}=-P_{\phi}=-\frac{\partial \mathcal{L}}{\partial \phi}=-\frac{2amr\sin^{2}\theta}{\rho^{2}}c\dot{t}+$$$$\frac{2a^{2}mr\sin^{2}\theta}{\rho^{2}}\sin^{2}\theta\dot{\phi}
\end{equation}
It is possible to show that impact parameter $\xi_{s}\equiv s \xi=\frac{sL_{z}}{E}$ [24, page 123]. Here, s defines the direction of motion of the light ray. For $s=+$, the light ray is in the direction of rotation of the mass and for $s=-$, the light ray is in the opposite direction of rotation [6]. Here $\xi_{s}$ is the first constant of motion. The second constant of motion is $\eta=\frac{\chi}{E^{2}}$, where $\chi= K-(L_{z}-aE)^{2}$. The quantity $\chi$ is known as the Carter constant [25] and $K$ is the conserved quantity along the direction $\theta$.
\\Following Carter [24, 25], we can have the four geodesic equations governing the motion of light ray in Kerr geometry as:
\begin{equation}
 \dot{\phi}=\frac{[2mra+(\rho^{2}-2mr)\xi\csc^{2}\theta]E}{(a^{2}+r^{2}-2mr)\rho^{2}}
\end{equation}
\begin{equation}
 c\dot{t}=\frac{[(r^{2}+a^{2})^{2}-\Delta a^{2}\sin^{2}\theta-2mra\xi]E}{\Delta \rho^{2}}
\end{equation}
\begin{equation}
 \dot{\theta}=\frac{(\eta+a^{2}\cos\theta-\xi_{s}^{2}\cot\theta)^{\frac{1}{2}}E}{\rho^{2}}
\end{equation}
\begin{equation}
  \dot{r}=\frac{[r^{4}-(\eta+\xi_{s}^{2}-a^{2})r^{2}+2mr(a-\xi_{s})^{2}+2mr\eta-a^{2}\eta]^{\frac{1}{2}}E}{\rho^{2}}
\end{equation}
 Now let us consider that the source of the light ray is not contained in the equatorial plane, but slightly above it by a distance $l$. Under this condition $\theta$ is not equal to $\frac{\pi}{2}$. If there is no gravitational field, the light ray will follow a straight line. The projection of this straight line on equatorial plane will be at a distance $u$ from the origin of the rotating mass (Fig 1), which we can call the projected impact parameter. At this minimum projected distance, the light ray has a height $l$ from the equatorial plane and $\psi$ is the angle that the light ray makes with the equatorial plane [14]. So,
\begin{equation}
 \tan\psi=\frac{l}{u}
\end{equation}
The value of $\psi$ is very small, as $l$ is only slightly above the equatorial plane. In other words $l$ is much smaller then $u$ ($l\ll u$).
\\These two parameters ($l, u$) can be used to level the light ray coming from infinity. The relation between the two parameters $(l, u)$ and two constants of motion ($\xi_{s}$ and $\eta$) can be written as [14] :
\begin{equation}
\xi_{s}=u\cos\psi
\end{equation}
and
\begin{equation}
 \eta=l^{2}\cos^{2}\psi +(u^{2}-a^{2})\sin^{2}\psi
\end{equation}
Now the relation between colatitude angle $\theta$ and the angle $\psi$ can be written as,
\begin{equation}
 \theta=\frac{\pi}{2}-\psi
\end{equation}
On the equatorial plane (i.e. $l=\psi=0$), $\xi_{s}=u$ and $\eta=0$.
\\Now, the geodesic equations can be written in terms of $\psi$ by using equation (15) in equation (8, 9, 10, 11) as follows:
$$\dot{\phi}=\frac{[2mr+(r^{2}+a^{2}\sin^{2}\psi-2mr)\xi_{s}\sec^{2}\psi]E}{(a^{2}+r^{2}-2mr)(r^{2}+a^{2}\sin^{2}\psi)}$$ Or,
\begin{equation}
 \dot{\phi}=\frac{[2mra+r(r-2m)\xi_{s} \sec^{2}\psi+a^{2}\xi_{s} \tan^{2}\psi]E}{(a^{2}+r^{2}-2mr)(r^{2}+a^{2}\sin^{2}\psi)}
\end{equation}
\begin{equation}
 \dot{\psi}=\frac{(\eta+a^{2}\sin\psi-\xi_{s}^{2}\tan\psi)^{\frac{1}{2}}E}{r^{2}+a^{2}\sin^{2}\psi}
\end{equation}
\begin{equation}
 c\dot{t}=\frac{[(r^{2}+a^{2})^{2}-\Delta a^{2}\cos^{2}\psi-2mra\xi_{s}]E}{\Delta(r^{2}+a^{2}\sin^{2}\psi)}
\end{equation}
\begin{equation}
  \dot{r}=\frac{[r^{4}-(\eta+\xi_{s}^{2}-a^{2})r^{2}+2mr(a-\xi_{s})^{2}+2mr\eta-a^{2}\eta]^{\frac{1}{2}}E}{r^{2}+a^{2}\sin^{2}\psi}
\end{equation}
 Here, we considered $\psi$ is very small i.e. $l\ll u$. So we can expand the trigonometric functions in the above geodesic equations up to second order terms as:
\begin{equation}
\tan^{2}\psi\simeq\sin^{2}\psi\simeq\psi^{2}
\end{equation}
and
\begin{equation}
\sec^{2}\psi\simeq1+\psi^{2}
\end{equation}
we can have the new format of geodesic equations by using equations (20), (21) in equations (16) to (19) as:
\begin{equation}
 \dot{\phi}=\frac{[2mra+r(r-2m)\xi_{s} (1+\psi^{2})+a^{2}\xi_{s} \psi^{2}]E}{(a^{2}+r^{2}-2mr)(r^{2}+a^{2}\psi^{2})}
\end{equation}
\begin{equation}
 \dot{\psi}=\frac{(\eta+a^{2}\psi-\xi_{s}^{2}\psi)^{\frac{1}{2}}E}{r^{2}+a^{2}\psi^{2}}
\end{equation}
\begin{equation}
 c\dot{t}=\frac{[(r^{2}+a^{2})^{2}-\Delta a^{2}(1-\psi^{2})-2mra\xi_{s}]E}{\Delta(r^{2}+a^{2}\psi^{2})}
\end{equation}
\begin{equation}
  \dot{r}=\frac{[r^{4}-(\eta+\xi_{s}^{2}-a^{2})r^{2}+2mr(a-\xi_{s})^{2}+2mr\eta-a^{2}\eta]^{\frac{1}{2}}E}{r^{2}+a^{2}\psi^{2}}
\end{equation}
In the present case, one can show that, r obtains a local extremum for the closest approach $r_{o}$, so that we can write:
 $$\dot{r}|_{r=r_{0}}=0$$
 It may be noted here that, $r_{0}=\sqrt{u^{2}+l^{2}}$. Further from equation (25), we can write :
 $$r_{0}^{4}-(\eta+\xi_{s}^{2}-a^{2})r_{0}^{2}+2mr_{0}(a-\xi_{s})^{2}+2mr_{0}\eta-a^{2}\eta=0$$
 or,
 \begin{equation}
 \frac{r_{0}^{2}}{\xi_{s}^{2}}=(1-\frac{a^{2}}{\xi_{s}^{2}})-\frac{2m}{r_{0}}(1-\frac{a}{\xi_{s}})^{2}+\frac{\eta}{\xi_{s}^{2}}$$
 $$-\frac{2m\eta}{r_{0}\xi_{s}^{2}}+\frac{\eta a^{2}}{\xi_{s}^{2}r_{0}^{2}}
 \end{equation}
\section{DEFLECTION ANGLE FOR OFF EQUATORIAL SOURCE}
Under the above kind of geometry, one may write the expression for the light deflection angle as : [26, page 189]
\begin{equation}
  \Delta\phi=2\int_{r_{0}}^{\infty}(\frac{d\phi}{dr})dr -\pi
\end{equation}
Using equation (22) and (25) in equation (27), one may write
\begin{widetext}
\begin{equation}
 \Delta\phi=2\int_{r_{0}}^{\infty}\frac{[2mra+r(r-2m)\xi_{s} (1+\psi^{2})+a^{2}\xi_{s} \psi^{2}]dr}{[a^{2}+r(r-2m)][r^{4}-(\eta+\xi_{s}^{2}-a^{2})r^{2}+2mr(a-\xi_{s})^{2}+2mr\eta-a^{2}\eta]^{\frac{1}{2}}}-\pi
\end{equation}
or,
\begin{equation}
 \Delta\phi=2\int_{r_{0}}^{\infty}\frac{\xi_{s}[\frac{2mra}{\xi_{s}}+r(r-2m)(1+\psi^{2})+a^{2}\psi^{2}]dr}
 {[a^{2}+r(r-2m)]\xi_{s}r^{2}[\frac{1}{\xi_{s}}-\frac{1-\frac{a^{2}}{\xi_{s}^{2}}}{r^{2}}+\frac{2m}{r^{3}}(1-\frac{a}{\xi_{s}})^{2}-\frac{\eta}{r^{2}\xi_{s}^{2}}+\frac{\eta}{r^{4}\xi_{s}^{2}}(2mr-a^{2})]^{\frac{1}{2}}}-\pi
\end{equation}
 As was done by some previous authors [15, 16], here we substitute $G=1-(\frac{a}{\xi_{s}})^{2}=1-\hat{a}^{2}(\frac{m}{\xi_{s}})^{2}$ and $F=1-(\frac{a}{\xi_{s}})=1-s\hat{a}\frac{m}{\xi_{s}}$, where $\hat{a}=\frac{a}{m}$. So the new form of the above deflection expression becomes:
\begin{equation}
 \Delta\phi=2\int_{r_{0}}^{\infty}\frac{[1-\frac{2m}{r}F+\psi^{2}(1-\frac{2m}{r}+\frac{a^{2}}{r^{2}})]dr}{[a^{2}+r(r-2m)]
 [\frac{1}{\xi_{s}^{2}}-\frac{G}{r^{2}}+\frac{2mF^{2}}{r^{3}}-\frac{\eta}{r^{2}\xi_{s}^{2}}+\frac{\eta}{r^{4}\xi_{s}^{2}}(2mr-a^{2})]^{\frac{1}{2}}}-\pi
\end{equation}
We substitute, $h=\frac{m}{r_{0}}$, $x=\frac{r_{0}}{r}$ and $n=\frac{\eta}{\xi_{s}^{2}}$. Accordingly we have,
$$dx=-\frac{r_{0}}{r^{2}}dr$$
and the limits will now change, when $r\longrightarrow \infty$, we have $x\longrightarrow 0$ and when $r\longrightarrow r_{0}$, then $x\longrightarrow 1$. Using these substitutions, in above equation we have:
\begin{equation}
\Delta\phi=2\int_{0}^{1}\frac{[1-2hxF+\psi^{2}(1-2hx+\hat{a}^{2}h^{2}x^{2})]dx}{[1-2hx+\hat{a}^{2}h^{2}x^{2}]
[\frac{r_{0}^{2}}{\xi_{s}^{2}}-Gx^{2}+2hF^{2}x^{3}-nx^{2}+2hnx^{3}-n\hat{a}^{2}h^{2}x^{4}]^{\frac{1}{2}}}-\pi
\end{equation}
Using the above substitutions, we can rewrite the equation (26) as :
\begin{equation}
\frac{r_{0}^{2}}{\xi_{s}^{2}}=(1-\frac{a^{2}}{\xi_{s}^{2}})-\frac{2m}{r_{0}}(1-\frac{a}{\xi_{s}})^{2}+\frac{\eta}{\xi_{s}^{2}}
 -\frac{2m\eta}{r_{0}\xi_{s}^{2}}+\frac{\eta a^{2}}{\xi_{s}^{2}r_{0}^{2}}$$
 $$=G-2hF^{2}+n-2hn+n\hat{a}^{2}h^{2}
\end{equation}
Combining equation (31) and (32), we can write

\begin{equation}
\Delta\phi=2\int_{0}^{1}\frac{[1-2hxF+\psi^{2}(1-2hx+\hat{a}^{2}h^{2}x^{2})]dx}{[1-2hx+\hat{a}^{2}h^{2}x^{2}]
[G(1-x^{2})-2hF^{2}(1-x^{3})+n(1-x^{2})-2hn(1-x^{3})+n\hat{a}^{2}h^{2}(1-x^{4})]^{\frac{1}{2}}}-\pi
\end{equation}

 or,

\begin{equation}
\Delta\phi=2\int_{0}^{1}\frac{[1-2hxF+\psi^{2}(1-2hx+\hat{a}^{2}h^{2}x^{2})]dx}{[1-2hx+\hat{a}^{2}h^{2}x^{2}]
[(G+n)(1-x^{2})-2h(F^{2}+n)(1-x^{3})+n\hat{a}^{2}h^{2}(1-x^{4})]^{\frac{1}{2}}}-\pi
\end{equation}
 Let us assume $G+n=g$ and $F^{2}+n=f$. If the light ray is contained in the equatorial plane, then we shall have $n=0$ and we have, $G=g$ and $f=F^{2}$. So the new form of the equation (34) will be,
\begin{equation}
\Delta\phi=2\int_{0}^{1}\frac{[1-2hxF+\psi^{2}(1-2hx+\hat{a}^{2}h^{2}x^{2})]dx}{[1-2hx+\hat{a}^{2}h^{2}x^{2}]
[g(1-x^{2})-2hf(1-x^{3})+n\hat{a}^{2}h^{2}(1-x^{4})]^{\frac{1}{2}}}-\pi
\end{equation}

or,

 $$\Delta\phi=2\int_{0}^{1}\frac{[1-2hxF+\psi^{2}(1-2hx+\hat{a}^{2}h^{2}x^{2})]dx}{[1-2hx+\hat{a}^{2}h^{2}x^{2}]
\sqrt{g(1-x^{2})}[1-2h\frac{f(1-x^{3})}{g(1-x^{2})}+\frac{n\hat{a}^{2}h^{2}}{g}(1+x^{2})]^{\frac{1}{2}}}-\pi$$
or,

 $$\Delta\phi=2\int_{0}^{1}\frac{[1-2hxF+\psi^{2}(1-2hx+\hat{a}^{2}h^{2}x^{2})]dx}{[1-2hx+\hat{a}^{2}h^{2}x^{2}]
\sqrt{g(1-x^{2})}\sqrt{1-2h\frac{f(1-x^{3})}{g(1-x^{2})}}[1+\frac{n\hat{a}^{2}h^{2}}{g}(1+x^{2})\{1-2h\frac{f(1-x^{3})}{g(1-x^{2})}\}^{-1}]^{\frac{1}{2}}}-\pi$$
 or,
$$\Delta\phi=2\int_{0}^{1}\frac{dx}{\sqrt{g(1-x^{2})}}[1-2hxF+\psi^{2}(1-2hx+\hat{a}^{2}h^{2}x^{2})][1-2hx+\hat{a}^{2}h^{2}x^{2}]^{-1}
\{1-2h\frac{f(1-x^{3})}{g(1-x^{2})}\}^{-\frac{1}{2}}$$
$$[1+\frac{n\hat{a}^{2}h^{2}}{g}(1+x^{2})\{1-2h\frac{f(1-x^{3})}{g(1-x^{2})}\}^{-1}]^{\frac{1}{2}}-\pi$$

For weak deflection limit, following [15, 16] one can assume, $m\ll r_{o}$, in other words, $h\ll 1$. So the above equation can be expanded in Taylor series in terms of $h$. By considering terms up to second order only, we can write :

\begin{equation}
 \Delta\phi=2\int_{0}^{1}\frac{dx}{\sqrt{g(1-x^{2})}}[1-2hxF+\psi^{2}(1-2hx+\hat{a}^{2}h^{2}x^{2})][1+2hx+x^{2}h^{2}(4-\hat{a}^{2})]
 $$$$[1+\frac{fh}{g}(\frac{1-x^{3}}{1-x^{2}})+\frac{3}{2}\frac{f^{2}h^{2}}{g^{2}}(\frac{1-x^{3}}{1-x^{2}})^{2}][1-\frac{n\hat{a}^{2}h^{2}}{2g}(1+x^{2})]-\pi
\end{equation}

Multiplying term by term and retaining only up to second order of $h$ we get,

\begin{equation}
 \Delta\phi=2\int_{0}^{1}\frac{dx}{\sqrt{g(1-x^{2})}}[1+h\{2x(1-F)+\frac{f}{g}\frac{(1-x^{3})}{(1-x^{2})}(1+\psi^{2})\}+\psi^{2}
 $$$$+h^{2}\{x^{2}(4-\hat{a}^{2}-4F)+\frac{2fx(1-F)}{g}\frac{(1-x^{3})}{(1-x^{2})}+\frac{3f^{2}}{2g^{2}}\frac{(1-x^{3})^{2}}{(1-x^{2})^{2}}(1+\psi^{2})
 -\frac{n\hat{a}^{2}}{2g}(1+x^{2})(1+\psi^{2})\}]-\pi
\end{equation}
Now integrating term by term and retaining terms only up to second order in $h$, we can write :
\begin{equation}
\Delta\phi=\pi(\frac{1+\psi^{2}}{\sqrt{g}}-1)+4h[\frac{f(1+\psi^{2})+g-Fg}{g^{\frac{3}{2}}}]+h^{2}[-4\frac{f}{g^{\frac{5}{2}}}\{f(1+\psi^{2})+g-Fg\}$$$$
+\frac{15\pi}{4}\frac{1}{15g^{\frac{5}{2}}}\{15f^{2}(1+\psi^{2})-4g(F-1)(3f+2g)-2g^{2}\hat{a}^{2}\}-\frac{3\pi n\hat{a}^{2}}{4g^{\frac{3}{2}}}(1+\psi^{2})]
\end{equation}
\end{widetext}
The above equation (38) represents the off-equatorial deflection of light due to a rotating mass in weak field limit. This expression is a function of mass, rotation parameter and $\psi$ only. We verify our result under some limiting conditions below:
\\If the light ray is contained in equatorial plane only, then $\psi$, $n$ will be equal to zero and $g=G$, $f=F^{2}$. Under these conditions we find, the above equation (38) will reduce to the following,
\begin{equation}
 \Delta\phi=\pi(\frac{1}{\sqrt{G}}-1)+4h[\frac{F+G-FG}{G^{\frac{3}{2}}}]+h^{2}[-4\frac{F(F+G-GF)}{G^{\frac{5}{2}}}$$
 $$+\frac{15\pi}{4}\frac{1}{15G^{\frac{5}{2}}}\{15F^{2}-4G(F-1)(3F+2G)-2G^{2}\hat{a}^{2}\}]
\end{equation}
 The above equation (39) is same as the expression for  equatorial deflection of light by Kerr mass obtained by Aazami et al. [16, equation no. (B17)].
\section{DISCUSSION AND CONCLUSION}
The main focus of the paper is to study the light deflection angle ($\Delta\phi$) as a function of the height $(l)$ of the light source above the equatorial plane in Kerr geometry. As obtained in equation (38), the deflection angle ($\Delta\phi$) is a function of $\psi\simeq \frac{l}{u}$, where $u$ is the projected impact parameter. So by varying the source height $(l)$ and projected impact parameter $(u)$, the main purpose of the paper can be achieved, to study their effect on the deflection angle.
\\In order to study the relation between $\Delta\phi$ and $l$ we obtain the values of $\Delta\phi$ as a function of $\tan\psi=\frac{l}{u}$ in Table1, by considering, the radius of Sun $6.955\times10^{8}$ meter as the closest approach of light ray ($r_{0}$) for both prograde and retrograde motions. The same was repeated, by considering a slow rotating pulsar PSR J 1748-2446 [27] for prograde and retrograde orbits and the results are reproduced in Fig.2. For these calculations, for the pulsar the equatorial projected impact parameter $(u)$ is taken to be $u=7500\times2m, 7501\times2m, 7502\times2m$ to maintain the weak field approximation. Subsequently, to understand the relation between $\Delta\phi$ and $u$, the plots were repeated in all the above mentioned cases, for three different values of $u/2m$. Here we should mention that, the pulsar has been chosen completely arbitrarily, except that the ratio of Kerr parameter(a) to Schwarzschild radius(2m) is same as that of Sun approximately, which will help us to compare results later. For that reason we choose a realistic pulsar PSR J 1748-2446.
\\In Fig2, we draw the deflection values ($\Delta\phi$) against $\tan\psi$ for the two cases (prograde and retrograde) for different parametric values of $u/2m$.
\\By studying the nature of the graphs, it is clear that the light deflection angle increases with the $\psi(\simeq\frac{l}{u}$) for both prograge and retrograde motion. For Sun (Table1) and pulsar (Fig2) the above variation of $\Delta\phi$ as a function of $\psi$ is clearly seen. On the other hand, it is quite obvious that the deflection angle decreases with the increase of projected impact parameter normalized by Schwarzschild radius ($u/2m$) (cf. Fig2 and Table1).
\\All the above mentioned calculations were done by considering $\psi$ to be very small $(l\ll u)$. Since from our expression we know that $\Delta\phi$ is a function of both $r$ and $\psi$, therefore if we consider $l$ to be in the same order as $u$, then instead of $\psi$ being a constant, it would be a function of $r$ itself and then we have to write the expression of light deflection angle as:
$$\Delta\phi=2\int_{r_{o}}^{\infty}[\frac{d\varphi\{r, \psi(r)\}}{dr}].dr -\pi$$
or,
$$\Delta\phi=2\int_{r_{o}}^{\infty}[\frac{\partial\phi}{\partial r}+\frac{\partial\phi}{\partial\psi}.\frac{\partial\psi}{\partial r}].dr-\pi$$

We may try to deduce the the values of $\frac{\partial\phi}{\partial\psi}$ and $\frac{\partial\psi}{\partial r}$, utilizing equation (16), (17) and (19). However, this we postpone for our future work and in the present work we report our work under the condition of $l\ll u$ only.

\begin{acknowledgments}
We are thankful to funding from UGC-SAP (DRS-1), which helped to complete the work and Dr.  A. Deshamukhya, Head Dept. of Physics, Assam University, Silchar, India for her encouragement to carry out this work. Finally, we are thankful to the anonymous referee of this paper for his/her very useful comments.
\end{acknowledgments}


\begin{table*}
\caption{\label{tab:table3} Variation of bending angle of light ($\Delta\phi$) as a function of $\psi\simeq\frac{l}{u}$ for three different values of $u/2m$ with $\hat{a}=0.5$ for Sun. To note, at the location of solar radius, value of $u/2m=2.35\times 10^{5}$ at $\psi=0$.}
\begin{ruledtabular}
\begin{tabular}{cccc}
 u/2m&$\psi\simeq\frac{l}{u}$&bending angle for prograde motion (arcsec)
&bending angle for retrograde motion (arcsec)\\\\ \hline
 $2.35\times 10^{5}$&$1\times10^{-5}$&1.753929&1.753933 \\
 &$2\times10^{-5}$&1.754125&$1.754129$\\
 &$3\times10^{-5}$&1.754451&1.754455\\
 &$4\times10^{-5}$&1.754906&1.754910\\\\ \hline
 $2.82\times10^{5}$&$1\times10^{-5}$ &1.461618&1.461621\\
 &$2\times10^{-5}$&1.461814&1.461817\\
 &$3\times10^{-5}$&1.462140&1.462142\\
 &$4\times10^{-5}$&1.462595&1.462597\\\\ \hline
 $3.29\times10^{5}$&$1\times10^{-5}$&1.252825&1.252826\\
 &$2\times10^{-5}$&1.253021&1.253023\\
 &$3\times10^{-5}$&1.253346&1.253803\\
 &$4\times10^{-5}$&1.253801&1.253803\\
\end{tabular}
\end{ruledtabular}
\end{table*}

 \begin{figure}
 \includegraphics[width=16cm]{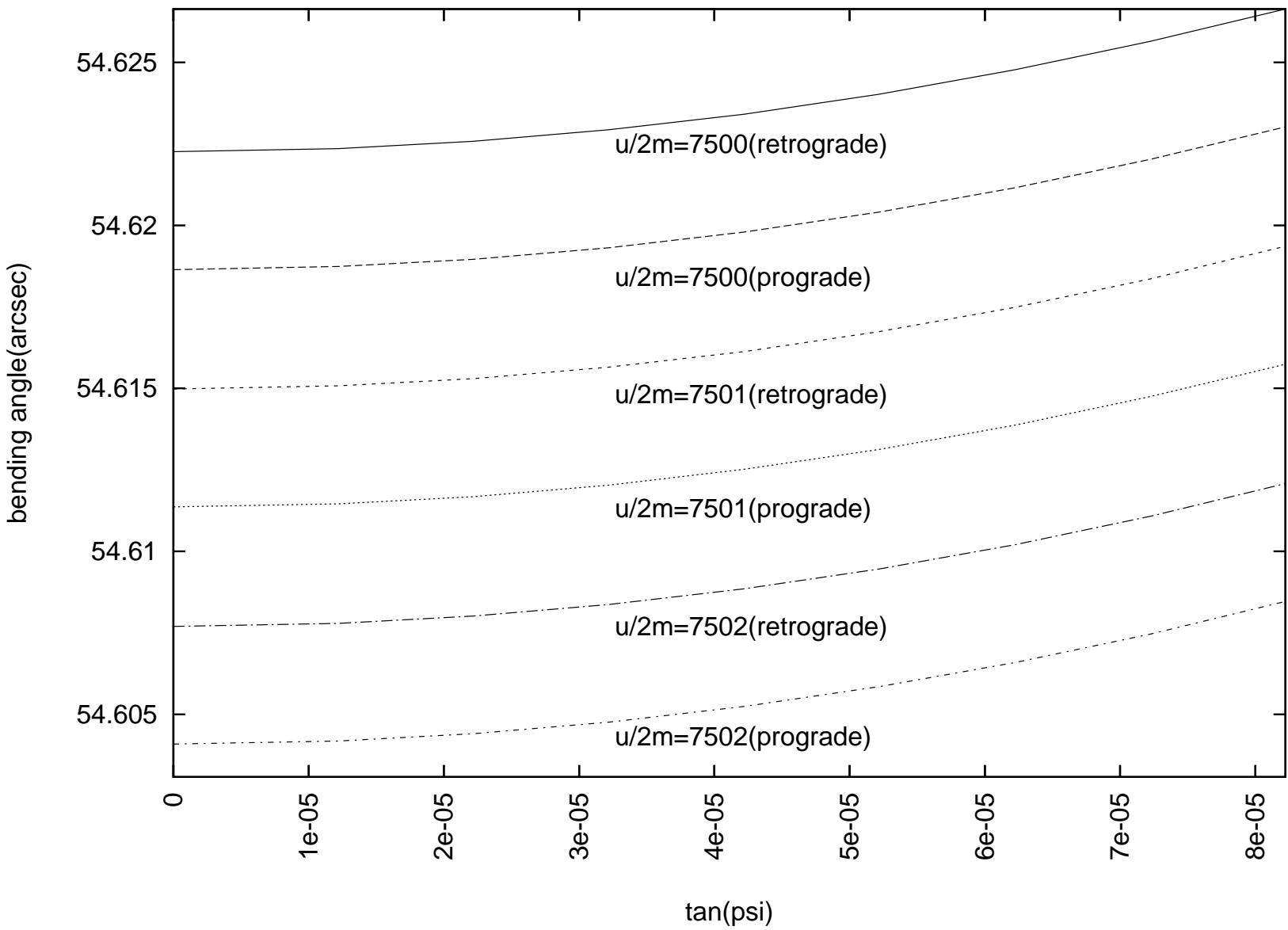}%
 \caption{\label{}Bending angle (arcsec) as a function of  $\tan\psi=\frac{l}{u}$ with constant rotation ($\hat{a}=0.5$) for both prograde and retrograde motion of PSR J 1748-2446 for different values of projected impact parameter (u/2m).}

 \end{figure}
\end{document}